\documentclass[fleqn,usenatbib]{mnras}

\usepackage{newtxtext,newtxmath}

\usepackage[T1]{fontenc}

\DeclareRobustCommand{\VAN}[3]{#2}
\let\VANthebibliography\thebibliography
\def\thebibliography{\DeclareRobustCommand{\VAN}[3]{##3}\VANthebibliography}

\usepackage{graphicx}	
\usepackage{amsmath}	
\usepackage{color}
\usepackage{enumitem}
\usepackage{hyperref}
\usepackage{verbatim}
\usepackage{mathrsfs}
\usepackage[all]{hypcap} 
\usepackage{afterpage}    
\usepackage{marginnote}
\usepackage{multirow}
\usepackage{lineno}
\usepackage{float}
\usepackage{siunitx}

\title[Magnetic fields in the galaxy NGC 3627]{Multi-Phase Magnetic Fields in the Galaxy NGC 3627}

\author[Liu et al.]{
Mingrui Liu$^{1}$,
Yue Hu$^{2,3}$\thanks{E-mail: yue.hu@wisc.edu},
A. Lazarian$^{3,4}$\thanks{E-mail: alazarian@facstaff.wisc.edu},
Siyao Xu$^{5}$,
Marian Soida$^{6}$\\
$^{1}$Department of Applied Physics and Applied Mathematics, Columbia University, New York, NY, 10076, USA\\
$^{2}$Department of Physics, University of Wisconsin-Madison, Madison, WI, 53706, USA\\
$^{3}$Department of Astronomy, University of Wisconsin-Madison, Madison, WI, 53706, USA\\
$^{4}$Centro de Investigación en Astronomía, Universidad Bernardo O’Higgins, Santiago, General Gana 1760, 8370993, Chile\\
$^{5}$Institute for Advanced Study, 1 Einstein Drive, Princeton, NJ 08540, USA (Hubble Fellow)\\
$^{6}$Obserwatorium Astronomiczne Uniwersytetu Jagiellonskiego, ul. Orla 171, 30-244 Kraków, Poland
}

\date{Accepted XXX. Received YYY; in original form ZZZ}

\pubyear{2022}

\begin{document}
\label{firstpage}
\pagerange{\pageref{firstpage}--\pageref{lastpage}}
\maketitle

\begin{abstract}
Magnetic fields play an important role in the formation and evolution of a galaxy, but it is challenging to measure them by observation. Here we study the Seyfert galaxy NGC 3627's magnetic field orientations measured from the synchrotron polarization observed with the Very Large Array (VLA) and from the Velocity Gradient Technique (VGT) using spectroscopic data. The latter employs the magnetohydrodynamical (MHD) turbulence's anisotropy to probe the magnetic fields. Being applied to the CO (2-1) and H$\alpha$ emission lines obtained from the PHANGS-ALMA and PHANGS-MUSE surveys, it reveals the magnetic field orientation globally consistent with the polarization. The agreement of the VGT-CO and polarization suggests that the magnetic fields associated with synchrotron emission also percolate through star-forming regions. The VGT-H$\alpha$ measurement reveals the magnetic fields in the warm ionized medium that permeates the disk and disk's vicinity so that it exhibits less agreement with polarization. We find prominent radial fields measured by synchrotron polarization appear in the transition regions from the spiral arms to the galactic bar, while such morphology is less apparent in the VGT-CO and VGT-H$\alpha$ measured magnetic fields. The radial fields suggest that the magnetic torque is important in removing orbiting gas' angular momentum. We notice that magnetic fields inferred from the dust polarization, VGT-CO, and synchrotron polarization are different in the east arm. We interpret this difference as arising from the fact that the three measurements are tracing the magnetic fields associated with pre-collision, the mixture of pre-collision and post-collision, and post-collision flows, respectively.
\end{abstract}

\begin{keywords}
general---galaxy---ISM: magnetic field
---turbulence---(magnetohydrodynamics)MHD
\end{keywords}



\section{Introduction}
Magnetic fields are pervasive in galaxies \citep{2005A&A...444..739B,2011MNRAS.412.2396F,2022arXiv220413611L,2022arXiv220605423H} and play a vital role in a number of important astrophysical processes. For instance, magnetic fields can provide supports against gravitational collapse (\citealt{1981MNRAS.194..809L,1987ARA&A..25...23S,2004RvMP...76..125M,2007ARA&A..45..565M,2012ARA&A..50...29C,2012ApJ...761..156F}) and regulate cosmic rays acceleration and transport (\citealt{1978MNRAS.182..147B,1978ApJ...221L..29B,1999ApJ...520..204G,2002ARA&A..40..319C,2004MNRAS.353..550B,2013ApJ...779..140X,2014ApJ...784...38L,2022arXiv220605423H}). Magnetic fields are also believed to be able to exert strong torques providing an efficient means to transport gas from a nuclear ring inward to feed an active galactic nuclei (AGN; \citealt{1984RvMP...56..255B,1989agna.book.....O,2012ApJ...747...60K,2012ApJ...751..124K}). Studying the role of magnetic fields is therefore essential for understanding galaxy evolution, especially how potential funneling gas can power Seyfert activities \citep{1996A&ARv...7..289M,1997ApJ...485..552M,2004ApJ...613..682P}.

In the past decades, our understanding of the plane-of-the-sky (POS) galactic magnetic field has been significantly advanced by measurements such as dust polarization \citep{2020ApJ...888...66L, 2021ApJ...923..150L,2022arXiv220413611L} and synchrotron polarization \citep{2005A&A...444..739B,2011MNRAS.412.2396F,2017A&A...600A...6M,2019A&A...623A..33S,2020A&A...639A.111S}. However, these techniques have limitations, while dealing with the complex multi-phase interstellar medium. The measurements of magnetic fields by any of the techniques are biased towards a particular component of the ISM. Therefore, the synergy of different measurements is essential for understanding of the actual magnetic field structure. 

In addition, all techniques have their intrinsic limiations. For instance, the Faraday rotation modifies synchrotron polarization \citep{1996ApJ...458..194M, 2006ApJ...637L..33H,2015A&A...575A.118O,2016ApJ...818..178L, 2016ApJ...824..113X} distorting the actual Plane-of-sky distribution of magnetic orientations. For dust polarimetry the variations of dust alignment efficiency  \citep{2007ApJ...669L..77L,2015ARA&A..53..501A} through the studied object can be a serious issue. 
On the top of this, polarization measurements are not capable of separating the magnetic fields directly associated with molecular gas and atomic gas in multi-phase interstellar media, which are important in understanding galaxy evolution (\citealt{2003A&A...407..485G}).

The confirmations of turbulence's ubiquity in galaxies \citep{1995ApJ...443..209A,2010ApJ...710..853C,2020NatAs...4.1064H} and interstellar medium (ISM; \citealt{2009ApJ...693..250B,2022arXiv220500012H}) call for utilizing the properties of Magnetohydrodynamic (MHD) turbulence (see \citealt{2019tuma.book.....B}) for magnetic field studies. One of the most important properties of such turbulence is its scale-dependent anisotropy induced by magnetic fields. These properties 
inspire a novel method, i.e., the Velocity Gradient Technique (VGT; \citealt{2017ApJ...835...41G}; \citealt{2017ApJ...837L..24Y}; \citealt{2018ApJ...853...96L}; \citealt{2018MNRAS.480.1333H}), to trace galactic magnetic fields. The physical basis for the technique stems from the anisotropy properties of MHD turbulence (\citealt{1995ApJ...438..763G,1999ApJ...517..700L}). Due to turbulent reconnection (\citealt{1999ApJ...517..700L}) turbulent eddies are not resisted if they rotate in the direction perpendicular to the local magnetic field that percolates the eddies.\footnote{The alignment of eddies along the local rather than mean magnetic field is proven numerically (\citealt{2000ApJ...539..273C,2001ApJ...562..279L,2002ASPC..276..170C}).}  This makes the small eddies aligned as needles along the local magnetic field. Therefore, the gradient of turbulent velocity is perpendicular to the eddy rotational axis and thus the direction of local magnetic fields. In particular, the gradient's amplitude increases especially at small scales, overwhelming the contribution from galactic shears and outflows (\citealt{2018ApJ...853...96L}) and resulting in a complementary way to polarization to trace galactic magnetic fields.  

The VGT's ability in tracing molecular-gas-associated and atomic-gas-associated magnetic fields has been widely tested in numerical simulations \citep{2018ApJ...853...96L,2020ApJ...905..129H,2021ApJ...911...53H} and observations \citep{2019NatAs...3..776H,2020ApJ...888...96H,2021ApJ...912....2H,2020MNRAS.496.2868L,2022A&A...658A..90A,2022MNRAS.510.4952L,2022arXiv220512084T,2022arXiv220606717Z}. It has proven that the self-absorption \citep{2019ApJ...874...25G,2019ApJ...873...16H,2021MNRAS.502.1768H} and weak self-gravity (i.e., except gravitational collapse; \citealt{2020ApJ...897..123H}) have little effects on the VGT. The requirements for VGT's application in extra-galaxies include that the turbulence injection, which is expected to be $\sim$ 100 pc for spiral galaxies similar to the Milky Way (\citealt{2010ApJ...710..853C}), can be resolved in observation, as well as that ion and neutral are well coupled. Such high-resolution observations are available in the PHANGS-ALMA targeting CO (2-1) emission (\citealt{2021ApJS..257...43L}) and PHANGS-MUSE targeting H-$\alpha$ emission (\citealt{2022A&A...659A.191E}) surveys. 



In this work, we study the galaxy NGC 3627. NGC 3627 is a barred spiral galaxy with Seyfert 2-type nucleus locating 11 Mpc away at R.A. (J2000) = $11^{\rm h}20^{\rm m}15^{\rm s}$, Dec. (J2000) = $+12^\circ 59'30''$ in the Leo Triplet galaxy group, having a $57^\circ$ inclination with respect to the POS and a position angle of $173^\circ$ measured North through East (\citealt{2003A&A...411..361K,2003A&A...412...45P,2012A&A...544A.113W}). The galaxy is a popular object of study for several interesting characteristics. Previous studies (e.g. \citealt{1979ApJ...229...83H}) proposed that a tidal encounter between NGC 3627 and the neighbor galaxy NGC 3628 about 800 Myr ago caused intense compression of interstellar medium in the western region of NGC 3627, resulting in a significantly bright polarization emission in the western spiral arm \citep{2012A&A...544A.113W}. Meanwhile, in the southeastern end of the galactic bar, magnetic fields measured by synchrotron polarization decouple with the optical dust lane (\citealt{2001A&A...378...40S}). Based on X-ray emission, \cite{2012A&A...544A.113W} suggested that this unusual magnetic field configuration may be due to a collision between NGC 3627 and a companion dwarf galaxy a few tens of Myr ago.The features of this collision were also widely observed by several studies of the gas kinematics of the galaxy using H I \citep{1993ApJ...418..100Z}, H$\alpha$\citep{2003A&A...405...89C}, and CO \citep{2011ASPC..446..111D}. Because of these intriguing features, NGC 3627 has been observed by many spectroscopic measures including the high-resolution PHANGS surveys, which makes it a good target of the VGT. The VGT's application to NGC 3627 was firstly performed by \cite{2022arXiv220605423H}. That work focused on mapping the galaxy's magnetic fields in the cold molecular phase using CO (2-1) emission and comparing with the magnetic fields measured by HAWC+ dust polarization at 154 $\mu$m (see the Appendix~\ref{appendix.b}). The observed good agreement between the two measurements proves the VGT's validity and encourages us to extend to mapping the magnetic fields in warm-phase media using the PHANGS-MUSE H-$\alpha$ emission. The VGT measurements are compared with the magnetic fields inferred from synchrotron polarization emission observed with the Very Large Array (VLA; \citealt{2001A&A...378...40S}) at 8.46 GHz/3.5 cm. The alignment and misalignment of the VGT and VLA polarization can have very important implications on many physical processes, e.g., star formation, and cosmic ray propagation. 

The paper is structured as such: in Sec.~\ref{sec:obs}, we briefly describe the used observational data. In Sec.~\ref{sec:method}, the methodological details of this work, including the theoretical consideration and the VGT pipeline, are presented. In Sec.\ref{sec:data}, we present the VGT-measured magnetic fields map and the comparison with the VLA polarization. Discussions about the alignment and misalignment observed in the VGT and VLA results are given Sec.~\ref{sec:disc}. We summarize this work in Sec.~\ref{sec:con}.

\section{Observations}
\label{sec:obs}
\subsection{CO (2-1) and H$\alpha$ emission lines}
The CO (2-1) and H$\alpha$ emission lines data are, respectively, from the PHANGS-ALMA ( \citep{2021ApJS..255...19L,2021ApJS..257...43L} and PHANGS-MUSE \citep{2022A&A...659A.191E} observation surveys, which aims at comprehensively studying the star formation process in nearby extragalaxies down to scale $\sim100$~pc. The distributions of CO and H$\alpha$ in NGC 3627 observed in these surveys are consistent with previous studies (e.g. \citealt{2003A&A...405...89C}), indicating a kinematical warp in the outer regions and an inner ring-like structure likely caused by ultra-harmonic resonant orbits.

The PHANGS-ALMA survey maps the CO (2-1) emission line (noted as CO hereafter for simplicity) in NGC 3627 using the Atacama Large Millimeter/submillimeter Array (ALMA). The survey has an angular resolution of $1.63''$ ($\sim90$~pc physically) and a channel width of 2.5 km s$^{-1}$. The RMS brightness temperature noise level of observations on NGC 3627 is $\sim 0.17$ K \citep{2021ApJS..257...43L}.

The PHANGS-MUSE survey utilizes the Multi Unit Spectroscopic Explorer (MUSE) at the Very Large Telescope (VLT) to map the H$\alpha$ emission line in NGC 3627. The galaxy is observed with 8 "pointings" blocks placed to obtain an $2''$ overlap between any neighboring area with a $5 \times 5$ arcmin$^{2}$ ($1''^{2}$) field of view, resulting in a FWHM resolution of $\sim$ 1.05$''$ ($\sim60$~pc physically; \citealt{2022A&A...659A.191E}). 

\subsection{Synchrotron polarization}
The magnetic field map for comparing with the VGT-measured results is inferred from polarized synchrotron emission of NGC 3627. The synchrotron polarization was observed with the VLA in November 1997 at 8.46 GHz with a half-power beam width (HPBW) resolution of $11''$\citep{2001A&A...378...40S}. The VLA polarization data was combined with single dish Effelsberg polarization map. The magnetic field orientation $\phi_{\rm B}$ is inferred from the synchrotron polarization orientation, namely $\phi_{\rm B}=\phi+\pi/2$, where the polarization angle $\phi$ is defined as:
\begin{equation}
\begin{aligned}
    \phi&=\frac{1}{2}\arctan(U,Q),\\
\end{aligned}
\end{equation}
where $Q$, $U$ are the Stokes parameters. In this work the Faraday rotation in the polarization data was not corrected as it does not exceed $\sim$ 5$^\circ$ at this frequency \citep{2001A&A...378...40S}.

\begin{figure*}
\label{fig.bfields}
    \centering
    \includegraphics[width=1\linewidth]{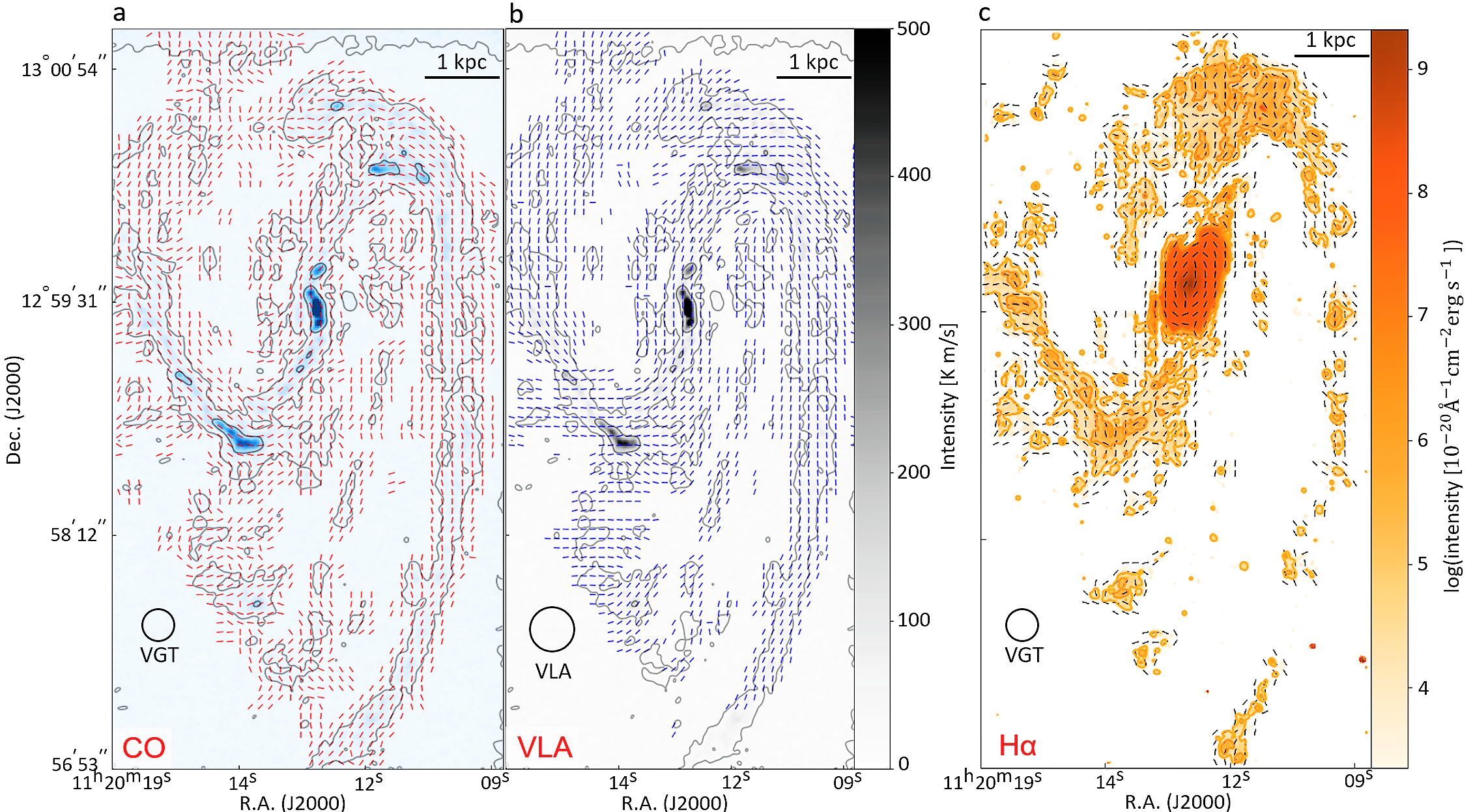}
    \caption{The magnetic field maps of NGC 3627. \textbf{Panel a}: probed by the VGT on the $\rm CO$ (2-1) emission line. \textbf{Panel b}: inferred from the VLA synchrotron polarization emission. \textbf{Panel c}: probed by the VGT on the H$\alpha$ emission line. The black circles approximate the resolutions of each observation. The black line segments on the top mark the scale of 1 kpc. The magnetic field measurements are laid upon NGC 3627's CO emission intensity map in Panel a, b, and the H$\alpha$ intensity map in Panel c.}
\end{figure*}

\section{Methodology}
\label{sec:method}
\subsection{Theoretical consideration of the VGT}
The theoretical basis of the VGT is the anisotropy of MHD turbulence. \cite{1995ApJ...438..763G}  introduced the critical balance condition: the cascading time ($k_{\perp}v_k)^{-1}$ equals the wave period ($k_{\parallel}v_{\rm A})^{-1}$. Here $k_{\perp}$ and $k_{\parallel}$ are the wavevectors perpendicular and parallel to the magnetic field direction, respectively. $v_{\rm A} $ denotes the Alfv\'en velocity and $v_k$ is turbulent velocity at scale $k$. \cite{1995ApJ...438..763G} derived the relation between the parallel and perpendicular wavenumbers: 
\begin{equation}
\label{eq.gs95}
\begin{aligned}
          \textit{$k_{\parallel}$}  \propto   \textit{$k_{\perp}^{2/3}$}.
\end{aligned}
\end{equation}

The original derivation of 
Eq.~\ref{eq.gs95} assumed that the $k_\|$ and $k_\bot$ are measured in respect to the mean magnetic field, which is not correct. The anisotropy measured with respect to the mean magnetic field is, in fact, dominated by the largest eddies of the cascade and is scale independent \citep{2000ApJ...539..273C,HXL21}. This is in striking contrast with the scale-dependent anisotropy given by Eq.~\ref{eq.gs95}.
Further research in \cite{1999ApJ...517..700L} showed that the anisotropy should be considered in respect to the local system of reference, i.e. in the system of reference that the turbulent motions take place. Due to fast turbulent reconnection, eddies with axis aligned with the local magnetic field direction undergo Kolmogorov-like cascade. This is possible as according to \cite{1999ApJ...517..700L} the time of the turbulent reconnection is equal to the eddy turnover time. In other words, the turbulent cascade follows the path of least resistance, i.e. induces turbulent eddies that minimally distort magnetic field. The critical balance in this interpretation is the equality of the period of the eddy turnover $l_\bot/v_{l,\bot}$ and the period of the Alfv\'en wave that is being induced this way, i.e. $l_\|/v_{\rm A}$. The Kolmogorov spectrum of the resulting Alfv\'enic turbulence follows automatically, as the cascade in the perpendicular direction is hydro-like. The importance of the local system of reference was demonstrated in \cite{2000ApJ...539..273C} and is used in all subsequent publications. 

\cite{1995ApJ...438..763G} assumed that the turbulence is being injected with $v_{\rm inj}=v_{\rm A}$, which means that the corresponding Alfven Mach number $M_{\rm A}=v_{\rm inj}/v_{\rm A}$ is equal to unity. This is a serious limitation. For $M_{\rm A}<1$, it was shown in 
\cite{1999ApJ...517..700L}
there exist a range of scales $[L_{\rm inj}M_{\rm A}^2, L_{\rm inj}]$ of weak turbulence with only the perpendicular scale changing with turbulent motions, i.e. $v_l\sim l_\bot^{1/2}$ and $l_\|$ being constant (see also \citealt{2000JPlPh..63..447G} for another derivation of the weak turbulence scaling). At the same time, for scales less that $L_{\rm inj} M_{\rm A}^2$, in the scales of strong MHD turbulence, the relation between the parallel and perpendicular scales of the eddies take the form \citep{1999ApJ...517..700L}:
 \begin{equation}
\begin{aligned}
l_{\parallel}  =  L_{\rm inj}(\frac{l_{\bot}}{L_{\rm inj}})^{2/3}M_{\rm A}^{-4/3},~M_{\rm A} \le 1.
\end{aligned}
\end{equation}

As the eddies move perpendicular to the local magnetic field, locally, they reveal the 3D direction of magnetic field similar as aligned dust grains reveal the magnetic field direction. Explicitly, in the range of strong MHD turbulence, the amplitude of velocity fluctuation and its corresponding gradient are \citep{1999ApJ...517..700L}:
\begin{equation}
\label{eq.4}
        \begin{aligned}
           &v_{l,\bot}  =  v_{\rm inj}(\frac{l_{\bot}}{L_{\rm inj}})^{1/3}M_{\rm A}^{1/3},~M_{\rm A} \le 1,\\
           &\nabla v_l\propto \frac{v_{l,\bot}}{l_{\bot}}=\frac{v_{\rm inj}}{L_{\rm inj}}M_{\rm A}^{1/3}(\frac{l_{\bot}}{L_{\rm inj}})^{-2/3},~ M_{\rm A} \le 1.
        \end{aligned}
\end{equation}
Notably, the gradient amplitude is proportional to $l_\bot^{-2/3}$. It implies the gradient amplitude increases at small scales. Such a property ensures the dominance of turbulent velocity's gradient at sufficiently small scales. In practice, the gradients can reveal the magnetic fields on the scales less than the turbulence injection scale $L_{\rm inj}$ which is $\sim 100$~pc for a typical spiral galaxy (see \citealt{2010ApJ...714.1398C}). The gradients arising from of galactic large scale shear get important on larger scales. Nevertheless, energetic outflows as well as regions of gravitational collapse can sometimes induce strong gradients that are comparable with those arising from turbulence. We consider this possibility in the paper. 


\subsection{The VGT pipeline}
To extract the velocity information in observation, the thin velocity channel Ch(x,y) of Doppler-shifted emission lines is employed. \cite{2000ApJ...537..720L} firstly proposed that due to difference in velocity, the structure of observed emission intensity in a given channel is significantly distorted. The narrower the channel width, the more significant distortion. When the channel width $\Delta v$ is smaller than the velocity dispersion $\sqrt{\delta (v^2)}$ of turbulent eddies under study, i.e., $\Delta v < \sqrt{\delta (v^2)}$, the intensity fluctuation in a thin channel is dominated by velocity fluctuation. Otherwise, the intensity fluctuation is dominated by density fluctuation in a thick channel. Therefore, the intensity gradient calculated in a thin velocity channel map (referred as velocity gradient) contains the information of velocity fluctuation \citep{2018ApJ...853...96L}. Note the observed fluctuation in a thin velocity channel map is anisotropic so it can be used to trace the POS magnetic field. However, the POS anisotropy can be affected by projection effect. When 3D magnetic field mainly orients along the LOS with little POS component, the POS anisotropy may disappear. Considering that NGC 3627 has inclination angle $\sim57$ degrees, the projection effect is insignificant globally.

The observed velocity dispersion of turbulence can rise up to $\sim10$ km/s at scale $\sim100$~pc in the Milky Way \citep{2022arXiv220500012H}. The CO emission data from the PHANGS-ALMA survey with a velocity resolution $\sim2.5$~km/s satisfies the criterion for a thin velocity channel. Moreover, for a cold medium, the thermal broadening effect is insignificant, so the condition $\Delta v < \sqrt{\delta (v^2)}$ accounts for only the turbulent velocity. However, such an effect cannot be ignored for warm medium H$\alpha$ in high temperatures ($\sim6000-10000$ K) due to the high sound speed $>10$ km/s. The velocity fluctuation's contribution in warm H$\alpha$ is suppressed relative to those in the cold medium when the velocity channels used are narrower than the warm H$\alpha$ thermal velocity. Therefore, the criterion for thin channel width should not be smaller than the sound speed. The H$\alpha$ emission data from the PHANGS-MUSE survey with velocity resolution $\sim35$~km/s satisfies this criterion. As velocity resolution is relatively lower than the one of CO emission, the contribution from density fluctuation increases but the contribution from velocity fluctuation does not vanish.

The velocity gradient of thin velocity channel is calculated as:
\begin{equation}
        \begin{aligned}
           \nabla_{x} {\rm Ch_i} (x,y) =  \textit{$G_{\rm x}$} * \textit{${\rm Ch_i} (x,y)$}, \\
           \nabla_{y} {\rm Ch_i} (x,y)  =  \textit{$G_{\rm y}$} * \textit{${\rm Ch_i} (x,y)$},
        \end{aligned}
        \end{equation}
where the asterisks indicate convolutions between the Sobel kernels\footnote{
\begin{equation}
       \begin{aligned}
                  &G_x = 
        \begin{pmatrix}
         -1 & 0 & +1\\ 
         -2 & 0 & +2\\
         -1 & 0 & +1
        \end{pmatrix},
        G_y =
        \begin{pmatrix}
         -1 & -2 & +1\\ 
        0 & 0 & 0\\
         +1 & +2 & +1
        \end{pmatrix}\\
        \end{aligned}
\end{equation}
} $G_{\rm x}$, $G_{\rm y}$ and totally $n_v$ thin channel maps ${\rm Ch_i} (x,y)$ with $i$ = 1, 2, ....., $n_v$. $\bigtriangledown_{x} {\rm Ch_i} (x,y)$, $\bigtriangledown_{y} {\rm Ch_i} (x,y)$ are the x, y gradient components in each thin channel map. The pixelized gradient map $\psi_{\rm g}^{\rm i} (x,y)$ is then calculated as:
\begin{equation}
       \begin{aligned}
&\psi{\rm ^i_g} (x,y)=\tan^{-1}(\frac{\nabla{}{}_{y} {\rm Ch_i} (x,y)}{\nabla_{x} {\rm Ch_i} (x,y)}).
\end{aligned}
\end{equation}

The resulting pixelized gradient map $\psi{\rm ^i_g}$ is processed by the sub-block averaging method \citep{2017ApJ...837L..24Y}. Velocity gradients within each equal-sized sub-block are employed to plot the gradient orientation histogram. The histogram is then fitted with a Gaussian distribution. The distribution's peak value gives the most probable orientation of gradients for each sub-block. The averaging produce is performed for every thin velocity channel. Using the averaged gradient angle $\psi{\rm ^i_{gs}}$, the pseudo-Stokes parameters are constructed as: 
\begin{equation}
\begin{aligned}
    &Q_{\rm g} (x,y)  =  \sum^{n_{\rm v}}_{{\rm i=1}} {\rm Ch_i} (x,y) \cos(2\psi{\rm ^i_{gs}} (x,y)),\\
    &U_{\rm g} (x,y)  =  \sum^{n_{\rm v}}_{{\rm i=1}} {\rm Ch_i} (x,y) \sin(2\psi{\rm ^i_{gs}} (x,y)),\\
    &\psi_{\rm g} = \frac{1}{2}\tan^{-1}(\frac{U_{\rm g}}{Q_{\rm g}}),
\end{aligned}
\end{equation}
where $n_{\rm v}$ is the total number of channels and $\psi_{\rm g}$ is the pseudo polarization angle. The direction of the POS magnetic field is thus defined as $\psi_{\rm B} = \psi_{\rm g} + \pi/2$ in the same manner as the synchrotron-inferred magnetic field orientation defined in Sec.~\ref{sec:obs}. 

\begin{figure}
\label{fig.am_dis}
    \centering
    \includegraphics[width=1\linewidth]{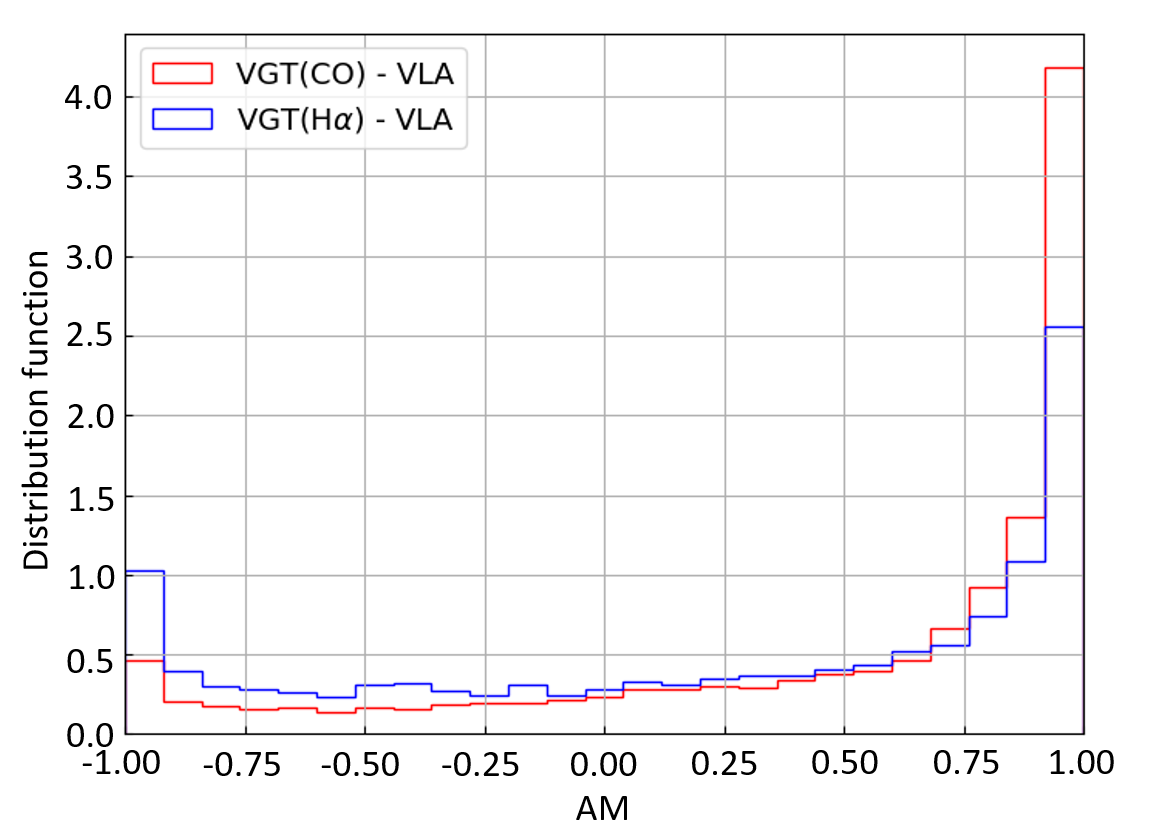}
    \caption{The histogram of AM for the VGT(CO)-VLA polarization and the VGT(H$\alpha$)-VLA polarization.}
\end{figure}

\begin{figure*}
\label{fig.am}
    \centering
    \includegraphics[width=1\linewidth]{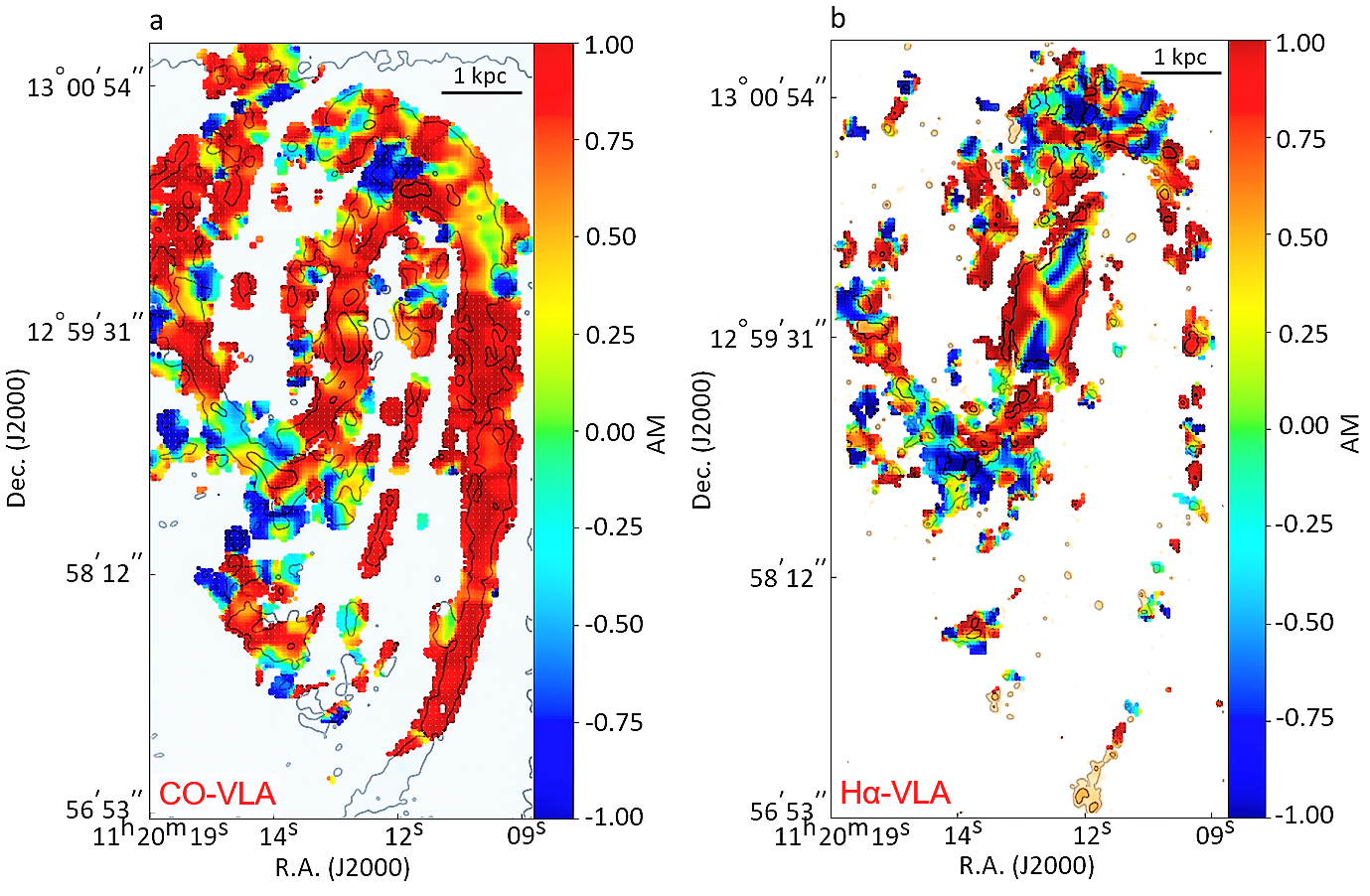}
    \caption{The spatial distributions of the AM. \textbf{Panel a}: AM between the VGT(CO) measurement and the VLA synchrotron polarization. \textbf{Panel b}: AM between the VGT(H$\alpha$) measurement and the VLA synchrotron polarization.}
\end{figure*}

\section{Results}
\label{sec:data}
\subsection{Magnetic alignment with galactic morphology}
Fig.~\ref{fig.bfields} presents the POS magnetic field maps inferred from the VGT using CO and H$\alpha$ emission lines, as well as the map obtained from the VLA synchrotron polarization. For the VGT's calculation, pixels with a brightness temperature less than three times the RMS noise are blanked out. The gradients are averaged as sub-blocks of 20 $\times$ 20 pixels. $\psi_g$ the resulting gradients map is smooth to $\rm FWHM\approx8''$.  The grid resolution of synchrotron emission data is 11$''$, close to the $\sim$ 8$''$ spatial resolution of the CO and H$\alpha$ emission data, which makes these measurements comparable with each other.

The CO emission exhibits clearly spiral structures, while the H$\alpha$ emission concentrated in the central nuclear region extending to the northern and southern transition regions from the spiral arms to the bar. In the two spiral arms, however, the H$\alpha$ emission is relatively faint. Moreover, magnetic field orientations measured by the VGT-CO generally follow the structure of the spiral arms. At the transition regions, the magnetic fields become aligned with the central bar. The magnetic fields inferred from synchrotron polarization exhibit similar morphology and generally agree with the VGT-CO measurement. It is noteworthy that near the southern end of the eastern spiral arm, the polarization-inferred magnetic field orientations do not align with the galactic structure indicated by CO but rather cross it. This pattern, however, is not observed in the VGT-CO measurement. As for the VGT-H$\alpha$ measurement, the corresponding map shows a less alignment with the synchrotron measurement.


\begin{figure}
\label{fig.am_d}
    \centering
    \includegraphics[width=1\linewidth]{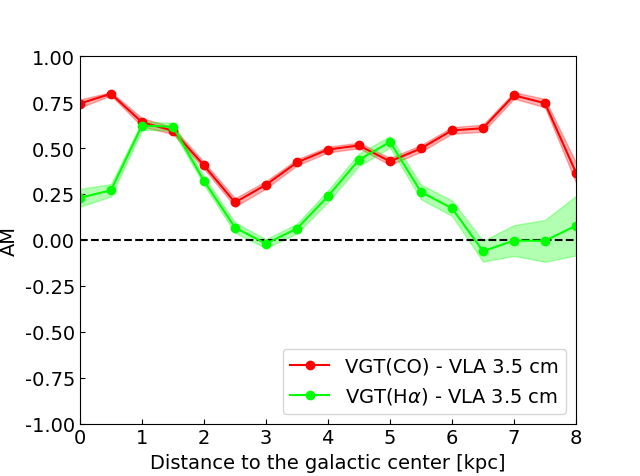}
    \caption{The average AM as a function of distance to the galactic center for the VGT(CO)-VLA polarization pair and the VGT(H$\alpha$)-VLA polarization pair. The shaded regions near the curves represent the sizes of uncertainty.}
\end{figure}

\begin{figure*}
\label{fig.pm}
    \centering
    \includegraphics[width=1\linewidth]{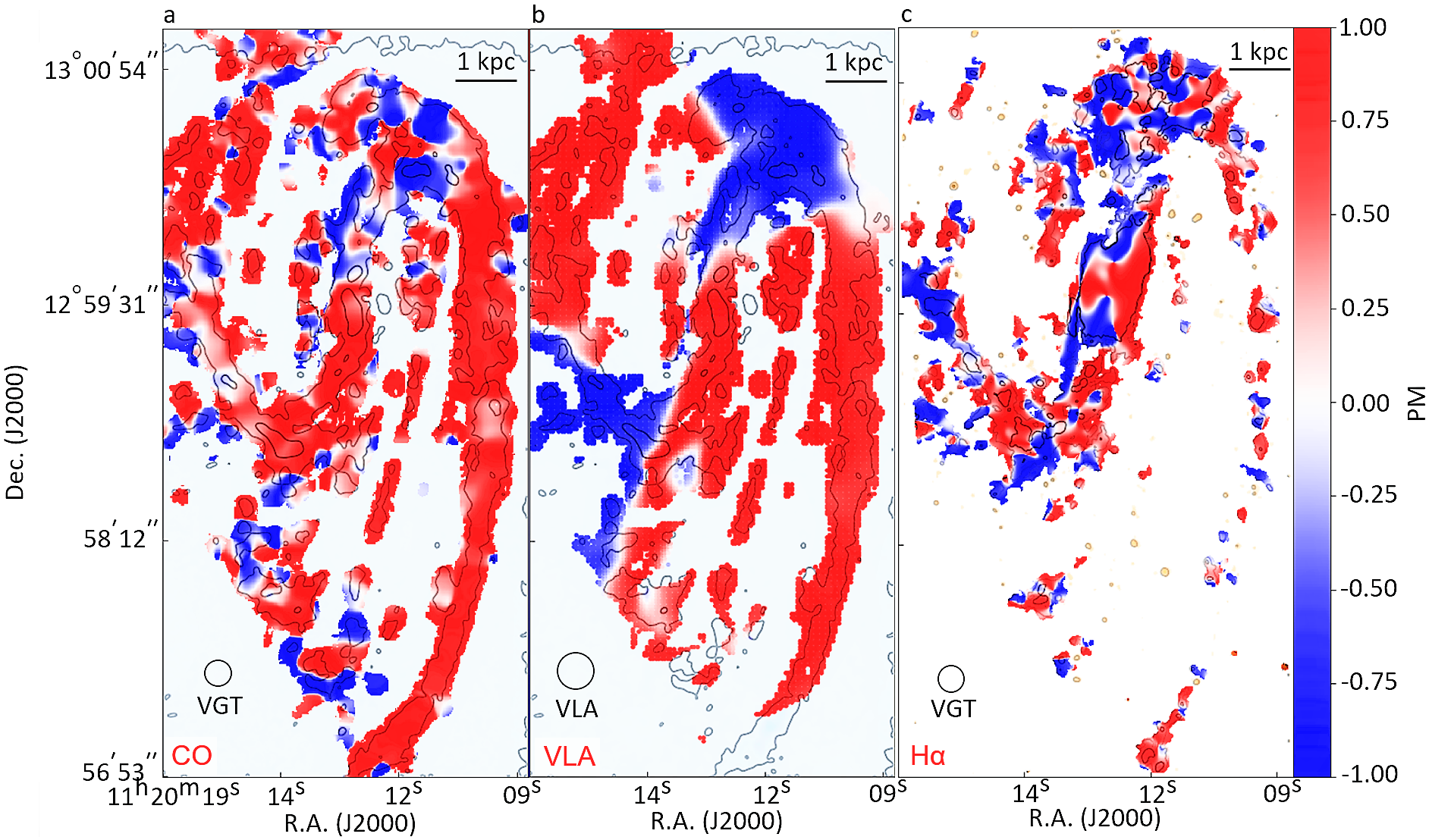}
    \caption{The spatial distributions of the PM. \textbf{Panel a}: PM of the VGT(CO) measurements. \textbf{Panel b}: PM of the VGT(H$\alpha$) measurements. \textbf{Panel c}: PM of the VLA synchrotron polarization.}
\end{figure*}

\subsection{Alignment and misalignment in multi-phase media}
The alignment of the VGT-CO, VGT-H$\alpha$, and synchrotron polarization measurements is quantified by the Alignment Measure (AM; \citealt{2017ApJ...835...41G}):
\begin{equation}
\begin{aligned}
    {\rm AM}  =  \cos(2\theta_{\rm r}),
\end{aligned}
\end{equation}
where $\theta_{\rm r}$ is the relative angle between the VGT-measured magnetic field vector $\phi_{\rm B}$ and the one $\psi_{\rm B}$ inferred from polarization. AM = 1 indicates that two vectors are parallel, while AM = -1 means that the two are orthogonal to each.

Fig.~\ref{fig.am_dis} shows the histograms of the AM. In general, although discrepancy (i.e., ${\rm AM<0}$) exists, the magnetic fields traced by the VGT-CO globally agree with the one inferred from polarization (i.e., ${\rm AM>0}$). 
As for the VGT-H$\alpha$ measurement, the fraction of positive AM decreases to only half of the VGT-CO's positive AM fraction, indicating less agreement with the polarization. Such alignments of the magnetic field measurements probed with VGT-CO and synchrotron polarization suggest that the magnetic fields associated with synchrotron emission percolate through molecular hydrogen at volume density $\approx10^{2}$~cm$^{-3}$ at least.



Fig.~\ref{fig.am} presents the AM maps of the VGT-CO, VGT-H$\alpha$, and synchrotron polarization measurement. An overall high AM > 0, i.e., good agreement, between the magnetic fields traced by the VGT-CO and synchrotron polarization is observed. This supports the correlation between star formation and CRs generation. Active star-forming regions are associated with a higher frequency of supernova explosions, which serve as efficient CR accelerators. CR electrons produced in star-forming regions are then diffuse to the low-density region, even to the galactic halos with much larger scale heights. The maximum diffused distance depends on the CR electrons' cooling time. The radiative cooling time of electrons responsible for a synchrotron photon of energy $\epsilon$ at frequency $f$ is $t_{\rm synch}\sim6.9\times10^6(\frac{B}{\rm 11\mu G})^{-1.5}(\frac{f}{{\rm 8.46 {\rm GHz}}})^{-0.5} {\rm years} \sim7$ Myrs, where the total magnetic field strength for NGC 3627 is $B\sim11 {\rm \mu G}$ \citep{2001A&A...378...40S}. This rapid cooling time is expected to company with a steep (non-thermal) flux density spectral index at high-frequency ranges. Such a steep index ($>0.8$) between 327 MHz and 1.4 GHz was observed in the spiral arms by \cite{2009A&A...503..747P}, while further work at higher frequency 8.46 GHz is required to confirm our expectation. By assuming a typical isotropic diffusion coefficient $\sim10^{28}{\rm cm^2/s}$  \citep{RevModPhys.48.161,1990acr..book.....B,1998ApJ...509..212S,2019PhRvD..99j3023E}, the diffusion lengths of CR electrons emitting synchrotron at 8.46 GHz and 1.4 GHz are $\sim500$~pc and $\sim1200$~pc, respectively. This agrees with the estimation given in \cite{2006ApJ...651L.111M}. They showed the CR electron's diffusion length scale at 1.4 GHz in NGC 3627 is $\sim$ 500~pc based on the far-infrared-radio correlation\footnote{The CRs' diffusion in star-formation regions and molecular clouds is suppressed due to the streaming of CRs along the tangled turbulent magnetic fields \citep{2022ApJ...927...94X} or mirror diffusion \citep{2021ApJ...923...53L,2021ApJ...922..264X}, resulting in a small diffusion coefficient.}.%
Therefore, CR electrons might travel only a few hundred parsecs from their acceleration sites and did not reach the typical scale height ($>0.2 - 1.0$~kpc) of abundant H$\alpha$ reservoir \citep{2018ApJ...862...25J}. This explains the good alignment between the VGT-CO measurements, which traces the magnetic field in star forming regions, and synchrotron polarization, because the CR electrons are still actively generated and well mixed with cold molecular gas, or confined in the vicinity ($<1$~kpc) of galactic disc. In addition, it is also possible that magnetic fields in star-forming regions around the disc are stronger than those at larger heights above the disc. CR electrons in this case might travel or diffuse further but synchrotron emission still is dominated by the contribution from strong magnetic fields. To confirm or exclude this possible explanation, observations of synchrotron polarization at lower frequencies are required. Either way, the VGT-CO and synchrotron polarization are tracing the magnetic fields in similar spatial regions.


Nevertheless, misalignment (AM < 0) between the VGT-CO and synchrotron polarization, as well as between the VGT-CO and dust polarization (see Appendix~\ref{appendix.b}), is notably observed in the transition region from the eastern arm to the central bar. Especially, the difference in the VGT-CO and synchrotron polarization is around $45^\circ$, while is $\sim90^\circ$ for the VGT-CO and dust polarization. This means that the three measurements give three different magnetic field morphology. Previous studies have recognized such interesting magnetic field inferred from synchrotron polarization near the eastern spiral arm \citep{2001A&A...378...40S}, where the magnetic fields discontinue to follow the morphological structure of the arm. The unusual magnetic field configuration may be explained by the eastern arm's collision with a dwarf galaxy 
\citep{1993ApJ...418..100Z,2003A&A...405...89C,2011ASPC..446..111D,2012A&A...544A.113W} taken place in a few tens of Myr ago. Due to CR electrons' short cooling time $\sim7$~Myrs, synchrotron generated before the collision has been aged. The detected synchrotron is more likely generated after the collision. Consequently, the synchrotron polarization at 8.46 GHz primarily traces the post-collision magnetic field. The time scale of tens of Myr, however, are not sufficient to generate a huge amount of post-collision dust. The dust polarization mainly trace the pre-collision magnetic field. As for CO, the collision generated new molecular gas and induced new gas flow. The pre-collision CO and post-collision CO get mixed so that the VGT-CO differs from the synchrotron polarization. Such a difference between the magnetic fields inferred from the three methods reflects the collision's effect in sharping the multi-phase ISM.

Fig.~\ref{fig.am_d} shows the AM averaged at linearly spaced annulus from the galaxy center (R.A. (J2000) = $11^{\rm h}20^{\rm m}15^{\rm s}$, Dec. (J2000) = $+12^\circ 59'30''$). The best agreement (AM > 0.5) of the VGT-CO and synchrotron polarization appears in approximately $r<2$~kpc and $4$~kpc $<r$, where $r$ is the distance from the center. The distance $2<r<4$~kpc corresponds to the transition region from the eastern arm to the central bar, and the AM decreases in that region as discussed above. As for the VGT-H$\alpha$, the AM values larger than 0.5 only appear at $1<r<2$~kpc and $r\sim5$~kpc. The former distance range corresponds to the central nuclear ring, while the latter indicates the western arm.

\subsection{Magnetic field morphology}
The magnetic field morphology imprints the motions of gas flow. A tangential, or spiral magnetic field is the result of galactic dynamo activities via differential rotations, while a radial field can be related to inflow or outflow. To characterize the magnetic field morphology, the pitch angle $\theta_p$ is calculated following the recipe proposed in \cite{2021ApJ...921..128B}. A zero pitch angle template is first generated by computing the radius $r$ and the zero pitch angle, which is perpendicular to the radial direction, of every pixel in galactocentric coordinates. Then the zero pitch angle and radius are transformed back to the observer’s coordinates, i.e., the POS. The pitch angle $\theta_{\rm pitch}$ is calculated as the difference between the measured position angle of the magnetic field and the template in the IAU convention, i.e., North through East. In addition, we defined the Pitch angle Measurement (PM; \citealt{2022arXiv220605423H}) as:
\begin{equation}
        \begin{aligned}
           {\rm PM}  =  \cos(2\theta_{\rm pitch}).
        \end{aligned}
\end{equation}
Similar to the AM, the PM is a scalar ranging from -1 to 1, which quantifies the physical morphology of magnetic fields. At locations where PM < 0, the corresponding magnetic fields are mostly radial; similarly, a positive PM value marks a place with a preferentially tangential magnetic field. PM = 0 indicates that the field is neither preferentially tangential nor radial.
\begin{figure*}
\label{fig.pm_d}
    \centering
    \includegraphics[width=1\linewidth]{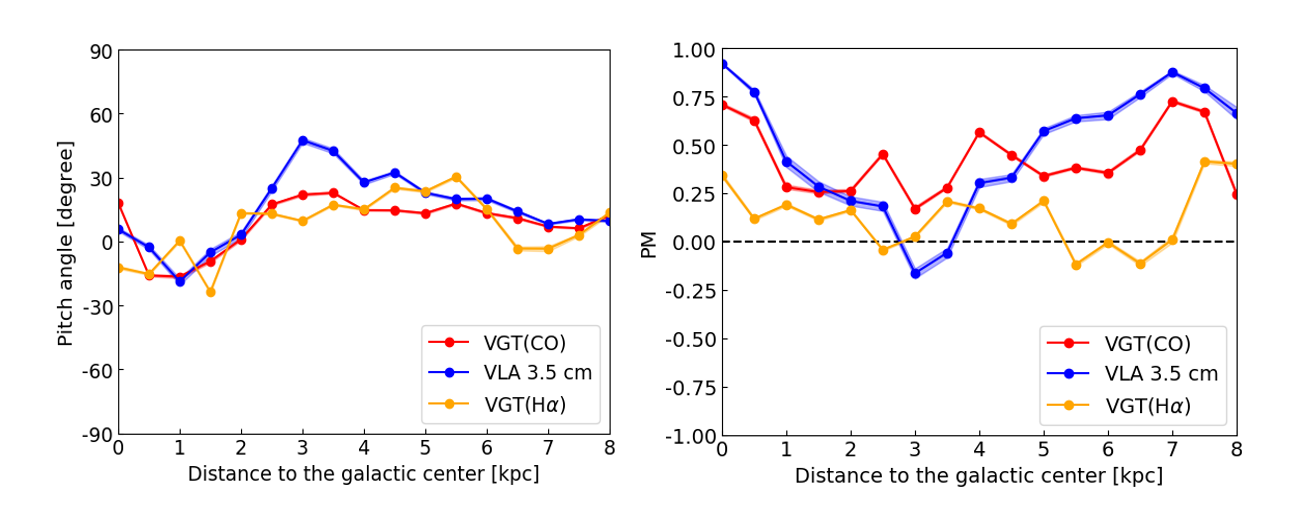}
    \caption{The average pitch angle (left panel) and PM (right panel) as a function of distance to the galactic center for the VGT(CO), VGT(H$\alpha$)-VLA, and VLA polarization measurements. The shaded regions near the curves represent the sizes of uncertainty.}
\end{figure*}

Fig.~\ref{fig.pm} presents the spatial distribution maps of the PM in the CO, H$\alpha$, and VLA synchrotron measurement. The PM map of synchrotron measurement indicates a spatially symmetric pattern. The magnetic fields are primarily tangential in a large area over the two spiral arms and the central nucleus area. In the transition regions between the galactic bar and the spiral arms, i.e., the southern end of the eastern arm and the northern end of the western arm, preferentially radial magnetic fields dominate. It is an important detail, as negative PM values coincides with on the positions of inflow towards the galactic nucleus \citep{2009ApJ...692.1623H,2017A&A...597A..85B}. The radial fields are therefore resulted by the gas inflow.  This agrees with the earlier numerical study performed by \cite{2012ApJ...751..124K}. In the transition regions, magnetic fields are compressed by shocks and stretched by shears in the dust lanes along the bar. The tension force of bent magnetic fields causes further removal of angular momentum of the gas at both dust-lane shocks. Such a loss in angular momentum is so large that the gas moving along the dust lanes makes a sharp turn toward the galaxy center. Parts of the inflow gas that further loses angular momentum are accreted into the central nucleus, but the rest encircles the center, gradually transforming into a nuclear ring. Consequently, a tangential field is again observed in the center.

The PM map of the VGT-CO exhibits a similar pattern to the synchrotron measurement, showing tangential fields in most galactic regions. The difference appears in the southern end of the eastern arm, where the magnetic fields induced from the VGT-CO are more tangential, but the magnetic fields inferred from synchrotron polarization are overall radial. This, however, supports the scenario that the pre-collision CR electron has cooled down and synchrotron polarization is tracing the magnetic fields associated with post-collision. The negative PM values in synchrotron polarization are symmetrically distributed in the southern end of the eastern arm and the northern end of the western arm, corresponding to the inflow gas losing angular momentum at dust-lane shocks. The intergalactic collision discussed before may generate new CO flow in the southern end of the eastern arm so that the negative PM values there are insignificant. A large portion of positive PM values in the VGT-CO is still observed in the northern end of the western arm, which is likely to be unaffected by the collision. On the other hand, the PM distribution of the VGT-H$\alpha$ presents a few distinctive patterns. It is more similar to the one of VGT-CO, but differences exist. This might be due to the fact that the warm gas phase H$\alpha$ is less sensitive to the Seyfert activity.
 
Fig.~\ref{fig.pm_d} shows the pitch angle and the PM averaged at each annulus from the galaxy center with linearly spaced radial bins. Notably, the pitch angles of all three measurements are $\sim-20^\circ$ when distance $r<2$~kpc, corresponding to the central nuclear ring. The angles change to $\sim20^\circ$ at $3<r<5$~kpc corresponding to the spiral arms. The sign change of the pitch angle suggests that the magnetic fields are outwardly oriented in the spiral arms, while the fields become inwardly oriented in the nuclear ring due to more substantial accretions. Unlike the pitch angle, the PM only quantifies whether the local field is preferentially radial or tangential. As shown in Fig.~\ref{fig.pm_d}, the average PM values of the VGT-CO and synchrotron polarization are both larger than 0.25 in the nuclear ring ($r<2$~kpc), suggesting preferentially tangential fields. The PM value of synchrotron polarization drops to negative at $r<3$~kpc, indicating radial fields in the spiral-arm-to-bar transition regions. At a larger distance, the PM value becomes positive again as the magnetic fields in spiral arms are preferentially tangential. The PM for the VGT-H$\alpha$ is slightly larger than 0 with some fluctuations.

\begin{table*}
\centering
\begin{tabular}{| c | c | c | c | c | c | c |}\hline
Emission & Spatial resolution ["] & Velocity resolution [km/s] & RMS noise & ${\rm \langle AM\rangle}$ & ${\rm \langle PM\rangle}$ & $\langle\theta_{\rm pitch}\rangle$\\\hline\hline
CO (2-1)$^{(2)}$ & $\sim$ 1.63 & 5 & $\sim$ 0.17 K & 0.50 & 0.37 & 10.45$^\circ$\\
H$\alpha$$^{(3)}$ & $\sim$ 1.05 & 35-80 & 7.92 $\times$ 10$^{-20}$ \AA$^{-1}$ cm$^{-2}$ erg s$^{-1}$  & 0.24 & 0.10 & 8.69$^\circ$\\
Synchrotron polarization$^{(1)}$ & $\sim$ 11 & - & 10 $\mu$Jy/beam & - & 0.40 & 20.66$^\circ$\\\hline
\end{tabular}
\caption{Important information of observations towards NGC 3627 used in this research. References: (1) Soida et al. (2001); (2) Leroy et al. (2021b); (3) Emsellem et al. (2022).
 }
\end{table*}

\section{Discussions}
\label{sec:disc}
\subsection{Prospects of the VGT}
In this work, an overall good agreement between the VGT-CO and synchrotron polarization measurements on the magnetic field orientations of NGC 3627 is presented. It adheres to the findings of \cite{2022arXiv220605423H}, the first application of the VGT on galactic magnetic fields, and confirms a good consistency between the VGT, synchrotron, and dust polarization measurement as well. These works have once again verified the reliability and accuracy of the VGT and introduced the application of such a novel technique to a broader context. It is also noteworthy that besides CO (2-1) and H$\alpha$ used in this work, the VGT can also be applied to other gas ISM like HCN, [Si II], [N II], etc. Such a wide range of available tracers introduces new opportunities for future applications of the VGT. Furthermore, the technique has the potential of decomposing galactic magnetic fields into components with different velocities as performed in \cite{2022MNRAS.510.4952L}. In galaxies, the rotation speeds of different parts such as spiral arms and nuclear rings are different. With the VGT that is capable of separating magnetic fields in each of these parts, it is now possible to directly investigate the magnetic properties of a specific galactic structure, which can thus provide novel details about the evolution of a galaxy.

\subsection{Implications from the agreement of synchrotron polarization and the VGT-CO measurement}
In this work, we observe the magnetic fields associated with the warm ionized synchrotron globally agree (i.e.,AM > 0.5) with the magnetic fields associated with the cold molecular gas CO, instead of the warm gas H$\alpha$. This suggests that the CR electrons in NGC 3627 are likely to be accelerated or produced in recent active star formations, which therefore have coupled magnetic fields with the local cold-phase CO gas instead of H$\alpha$ in the same warm diffuse phase. This finding agrees with the tight correlation between far-infrared and radio luminosities of star-forming galaxies \citep{1971A&A....15..110V,1973A&A....29..263V,2001ApJ...554..803Y}, which also shows the strong correlation between star formation and CRs. This also supports that high-energy CR electrons might travel only a few hundred parsecs and have not yet reached/diffused to the region at scale height ($>0.2 -1 .0$~kpc) due to the short cooling-time or suppressed diffusion in star-forming regions \citep{2022ApJ...927...94X,2021ApJ...923...53L,2021ApJ...922..264X}. Another possibility is that although an ionized (high-frequency) radio halo exits, as suggested by \cite{2011A&A...531A.127S} for the galaxy NGC 5775, star-forming regions are typically associated with strong magnetic fields and their contribution might dominates the synchrotron emission. To give a definite conclusion, radio observations at lower frequencies are required to compare with VGT-H$\alpha$.


\subsection{Implications from magnetic field morphology}
Quantified magnetic field morphology with the VGT in NGC 3627 can contribute to the study of the dynamo process in nearby galaxies. To summarize, the dynamo mechanism may lead to preferentially spiral magnetic fields that track the differential rotation shear (\citealt{1996ARA&A..34..155B}) at large scales. At scales smaller than turbulence's injection scale, turbulent dynamo amplifies magnetic fields \citep{2016ApJ...833..215X}. \cite{2022arXiv220605423H} discussed the dynamo mechanism concerning the magnetic field morphology of several Seyfert galaxies including NGC 3627 using dust polarization (see Appendix~\ref{appendix.b}). In the studied galaxies of that work, spiral magnetic fields are detected globally, while in the bar-outskirt transition regions, preferentially radial magnetic fields appear. The observed magnetic field morphology in NGC 3627 in this work adheres to the previous findings, except for the east arm region. A collision with a dwarf galaxy in that region likely causes each of the magnetic field measurements to trace different gas flow, as proposed in Sec.~\ref{sec:data}. 

In addition, radial magnetic fields are observed symmetrically in the two bar-spiral-arm transition regions in the synchrotron measurement, while in the VGT-CO measurement, the radial field components are less abundant in such regions, showing no similar symmetric distribution. Such morphological differences in the magnetic fields suggest a larger significance of the streaming motion of galactic gas, especially the new gas flow in the southeastern transition region introduced by the collision discussed before. This supports that the synchrotron polarization imprints the magnetic fields associated with the post-collision gas flow, while the VGT-CO is contributed by the post-collision and pre-collision gas flows. Moreover, the observed radial field in the transition regions agrees with the theory of Seyfert fueling activity that magnetic field torque helps the removal of the gas' angular momentum so that the inflow gas are accreted into the central nucleus
\citep{2012ApJ...751..124K}. 

\section{Conclusions}
\label{sec:con}
In this work, the magnetic field properties of the spiral galaxy NGC 3627 are studied using the VGT with observational data of the CO (2-1) emission line and the H$\alpha$ emission line. We compare the VGT measurements with the magnetic field measurement inferred from the existing VLA synchrotron polarization to understand the magnetic fields in multi-phase astrophysical media in NGC 3627. To summarize our main findings:
\begin{enumerate}
    \item The magnetic field orientations probed with the VGT on CO (2-1) emission line are generally consistent with the magnetic field configurations induced from the VLA synchrotron polarization. This once again testifies the reliability and accuracy of the VGT for probing interstellar/galactic magnetic fields, as well as demonstrates coherent magnetic field components in different interstellar phases. 
    \item The magnetic field orientations probed with the VGT-CO show a better alignment with the synchrotron polarization than the VGT-H$\alpha$. It suggests the magnetic fields in cold molecular CO gas percolate with that measured by 8.46 GHz synchrotron polarization in NGC 3627. This supports the strong spatial correlation between CO and synchrotron emission regions, suggesting CR electrons have a relatively short diffusion length or the synchrotron emission is dominated by potential strong magnetic fields in star-forming regions.  Together with the VGT-H$\alpha$, the three measurments reveals the magnetic field configuration permeating the disk and disk's vicinity.
    \item Major disagreements between the VGT-CO and synchrotron measurement are detected near the southern end of NGC 3627's eastern spiral arm, which is suspected to be a collision site of the galaxy with a dwarf galaxy \citep{1993ApJ...418..100Z,2003A&A...405...89C,2011ASPC..446..111D,2012A&A...544A.113W}, introducing new gas flow near this region. The synchrotron polarization imprints magnetic fields in the post-collision gas flow, while the VGT-CO traces magnetic fields associated with the mixture of post-collision and pre-collision flow, which potentially causes such a discrepancy.
    \item Synchrotron measurement on the magnetic field morphology of NGC 3627 suggests highly symmetrical radial magnetic fields in the bar-outskirt transition areas, while the VGT-CO possesses a larger abundance of tangential field components, especially near the southeastern collision region. This implies that the magnetic fields are important in removing the gas' angular momentum to fuel the Seyfert nucleus.
\end{enumerate}

\section*{Acknowledgements}

Y.H. acknowledges the support of the NASA TCAN 144AAG1967. A.L. acknowledges the support of the NASA ATP AAH7546. S.X. acknowledges the support for this work provided by NASA through the NASA Hubble Fellowship grant \# HST-HF2-51473.001-A awarded by the Space Telescope Science Institute, which is opearted by the Association of the Universities for Research in Astronomy, Incorporated, under NASA contract NAS5-26555.

\section*{Data Availability}
The data underlying this article will be shared on reasonable request to the corresponding author.



\bibliographystyle{mnras}
\bibliography{example} 




\begin{figure*}
\label{fig.uce}
    \centering
    \includegraphics[width=1\linewidth]{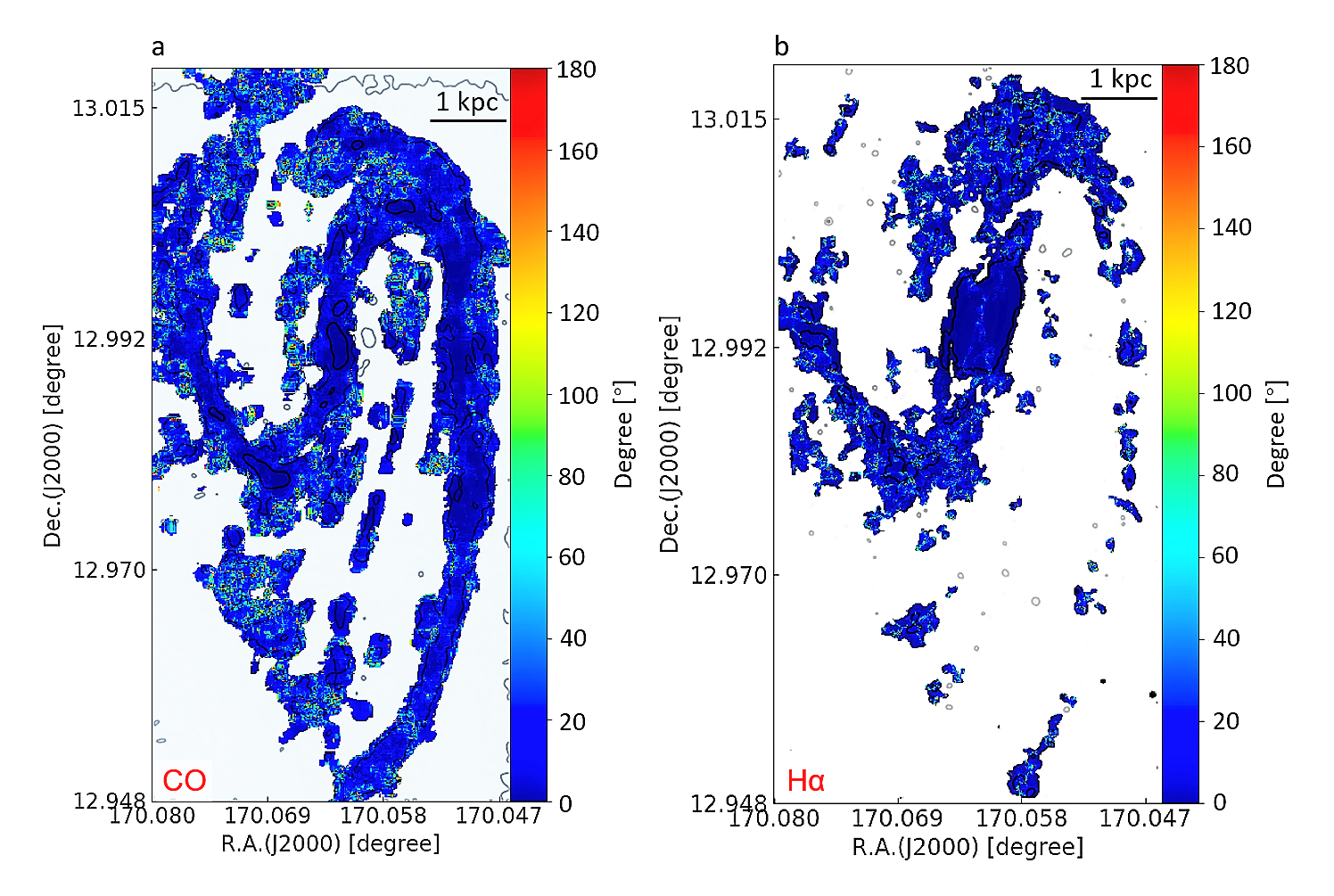}
    \caption{The uncertainty maps of VGT-CO, VGT-H$\alpha$ measurements.}
\end{figure*}

\appendix
\section{Uncertainty of magnetic field measurements}
Fig.~\ref{fig.uce} presents the uncertainty maps of our magnetic field measurements. Such uncertainties can be mostly attributed to the systematic error from the observational data as well as the VGT algorithm. As introduced in Sec.~\ref{sec:method}, the algorithm fits the gradient’s orientations over predetermined sub-regions into a Gaussian histogram and yields the angle at which exists the peak value of the histogram. This error from the Gaussian fitting algorithm within the 95\% confidence level is denoted as $\sigma_{\psi_{gs}} (x, y, v)$, and thus the uncertainties are obtained as such:
\begin{equation}
\begin{aligned}
    &\sigma_{{\rm cos}} (x,y,v)  =  |2 {\rm sin}(2 \psi_{\rm gs} (x,y,z)) \sigma_{{\psi_{\rm gs}}} (x,y,v)|\\
    &\sigma_{{\rm sin}} (x,y,v)  =  |2 {\rm cos}(2 \psi_{\rm gs} (x,y,z)) \sigma_{{\psi_{\rm gs}}} (x,y,v)|\\
    &\sigma_{q} (x,y,v)  =  |{\rm Ch} \cdot {\rm cos(2 \psi_{\rm gs})}|\sqrt{(\sigma_{n}/{\rm Ch})^2+(\sigma_{{\rm cos}}/{\rm cos(2 \psi_{\rm g})})^2}\\
    &\sigma_{u} (x,y,v)  =  |{\rm Ch} \cdot {\rm sin(2 \psi_{\rm gs})}|\sqrt{(\sigma_{n}/{\rm Ch})^2+(\sigma_{{\rm sin}}/{\rm sin(2 \psi_{\rm g})})^2}\\
    &\sigma_{Q} (x,y)  =  \sqrt{\sum_{v} {\sigma_{q}(x,y,v)^2}}\\
    &\sigma_{U} (x,y)  =  \sqrt{\sum_{v} {\sigma_{u}(x,y,v)^2}}\\
    &\sigma_{\psi_{g}} (x,y) = \frac{|U_{\rm g}/Q_{\rm g}|\sqrt{(\sigma_{Q}/Q_g)^2+(\sigma_{U}/U_g)^2}}{2[1+(U_{\rm g}/Q_{\rm g})^2]}
\end{aligned}
\end{equation}
where $\sigma_{\psi_g}$ is the angular uncertainty of the magnetic field measurements, $\sigma_n (x, y, v)$ denotes the noise in velocity channel Ch$(x, y, v)$ as well as error propagation, and $\sigma_Q (x, y)$, $\sigma_U(x, y)$ give the respective uncertainty of the pseudo-Stokes parameters $Q_g (x, y)$, $U_g (x, y)$. As shown in Fig.~\ref{fig.uce}, the VGT algorithm produces low-noise measurements with an uncertainty median of 11.50$^\circ$ for the VGT-CO and 5.33$^\circ$ for the VGT-H$\alpha$, respectively.
\section{Magnetic field inferred from dust polarization}
\label{appendix.b}
\begin{figure}
\label{fig.dust}
    \centering
    \includegraphics[width=1\linewidth]{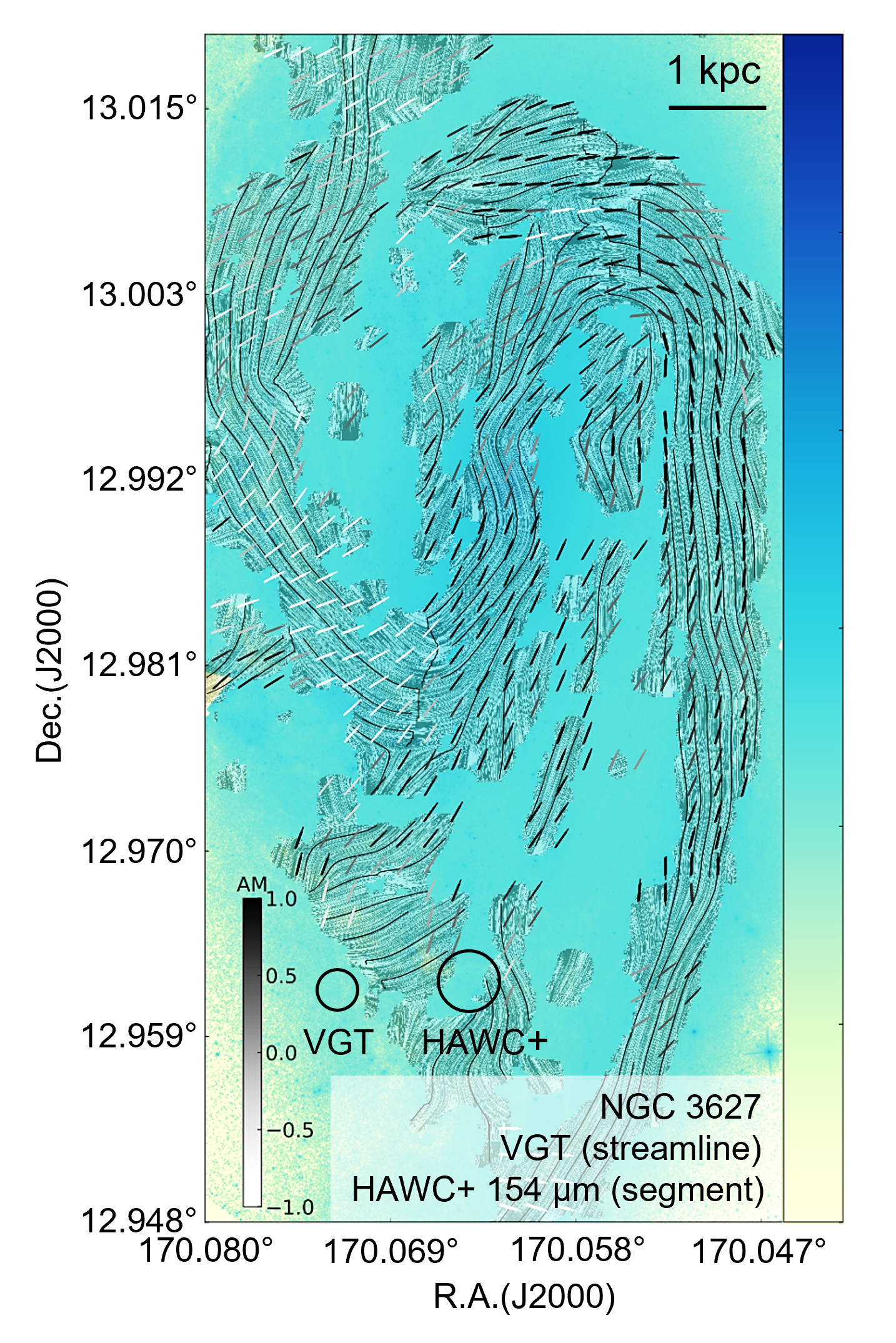}
    \caption{Morphology of magnetic fields revealed by the VGT using $\rm ^{12}CO$ (J = 2-1) emission line and HAWC+ polarization at 154 $\mu$m. The VGT-measurement is visualized by black streamlines and HAWC+ is represented by the colored segments. Colors on polarization vectors present the AM of the VGT and polarization. The colorbar of background HST WFC3/F814W ultraviolet image is logarithmically spaced in the range from $10^{-6}$ to $10^{6}$ electrons per second. Extracted from \color{blue}{Hu et al. (2022)}.}
\end{figure}

Fig.~\ref{fig.dust} presents the comparison of the magnetic field inferred from the VGT and HAWC+ dust polarization at 154 $\mu$m. The VGT and HAWC+ exhibit agreement in the west arm. However, in the east arm, where the collision with a dwarf galaxy happened, the VGT and HAWC+ are almost perpendicular to each, i.e., AM$\sim -1$. The result is extracted from \cite{2022arXiv220605423H}. 
 

\bsp	
\label{lastpage}
\end{document}